\documentclass[prb,twocolumn,showpacs]{revtex4}
\usepackage{amsmath,amssymb,graphicx,color,bm,epstopdf}

\begin{document}

\title{A complete architecture of integrated photonic circuits based on AND and NOT logic gates of exciton-polaritons in semiconductor microcavities}
\author{T. Espinosa-Ortega}
\affiliation{Division of Physics and Applied
Physics, Nanyang Technological University 637371, Singapore}

\author{T. C. H. Liew}
\affiliation{Division of Physics and Applied
Physics, Nanyang Technological University 637371, Singapore}

\begin{abstract}
We present a complete photonic logic gate architecture in a compact solid-state system, making use of the non-linear and spintronic properties of exciton-polaritons in semiconductor microcavities. The dynamics of the system is modelled using the spinor Gross-Pitaevskii equations and it is shown that the proposal fulfills all the necessary criteria for fully functioning information processing devices without the use of any external electric fields.
\end{abstract}

\maketitle

\section{Introduction}

The construction of photonic devices typically relies on electro-optical effects to provide functionality as in recent modulator designs based on ring resonators \cite{1}, graphene \cite{2} and exciton condensates \cite{3}. Fully photonic devices, that do not require external electric fields, are harder to come by due to the limited optical nonlinear response of typical materials. While on-chip diodes have been recently demonstrated \cite{4}, a complete photonic logic gate architecture has not yet been developed in a compact system, which would allow realization of optical microprocessors. Here we introduce such an architecture in a system based on the spintronic properties of exciton-polaritons \cite{5}. This hybrid light-matter design is highly compact and offers the advantages expected of photonic circuits including \cite{6}: no distance-dependent loss or signal gradation; electrical isolation and immunity to electromagnetic interference; reduced error-checking overhead; and direct coupling to external light, useful for fibre optic communication.

For a functional circuit architecture, five criteria must be fulfilled \cite{7}: $1)$ \textit{Universal logic} – both AND and NOT type gates (or an equivalent set) are required; $2)$ \textit{Cascadability} – the output of one gate must be able to drive the next; $3)$ \textit{Fan Out} – it must be possible to split and duplicate signals; $4)$ \textit{Amplification} – loss must be fully compensated such that signals are maintained at the logic-level; $5)$ \textit{Input-output isolation} – the circuit must operate in one direction only, with no significant feedback effects from the output.

Beginning with the first criterion, one requires some kind of non-linear element. In exciton-polariton systems -hybrid states of light confined in a microcavity and quantum well excitons \cite{5}- this is provided from exciton-exciton interactions in the form of an effective Kerr type nonlinearity. These nonlinear interactions have allowed the demonstration of optical modulators/transistors \cite{8,9} and amplifiers \cite{10,11}. Exciton-polaritons also carry a spin \cite{12}, which can take values $+1$ or $-1$ when projected on the structure growth axis, and this has allowed the construction of spin switches \cite{13,14}. Unfortunately, none of the aforementioned devices has demonstrated the criterion of \textit{Cascadability} or \textit{Fan Out} and external optics would likely be necessary to do so, making for bulky and unscalable devices. Very recently a scheme of Casadability and Fan-Out was demonstrated \cite{Dario}, based on earlier work on the controlled shifting of hysteresis curves \cite{Giorgi}. However, the scheme was not completed with a Universal Logic. Indeed, as one tries to fulfil more criteria simultaneously, the task becomes exponentially more challenging.

On the theoretical side, a design for optical circuits was proposed in $2008$ \cite{15}, making use of the phenomenon of spin-multistability \cite{16} in resonantly excited microcavities. Spin-polarized signals were carried by propagating spin-domains along channels known as polariton neurons. This theoretical proposal solved the problems of \textit{Cascadability}, \textit{Amplification} and \textit{Fan-Out}, since the spin-domain propagation continued without decay despite the finite lifetime of exciton-polaritons, which was particularly short at the time ($~3ps$). However, a \textit{universal set} of logic gates was not realized: it was possible to construct AND type or OR type logic gates (but not both simultaneously) and it was not known how to construct a NOT type logic gate. Both AND and NOT type gates are required for \textit{Universal Logic}.

More recently, improvements in sample growth technology have increased the exciton-polariton lifetime by orders of magnitude \cite{17} and the ballistic propagation of spin signals can proceed over hundreds of microns \cite{18} with minimal scattering. We make use of this development in combination with spin precession in 1D exciton-polariton waveguides \cite{19} to couple multiple nodes in a network designed for the realization of optical circuits (see Figure 1). By exciting each node in the network by a continuous-wave (cw) pump, the presence of polarization multistability allows each node to be switched between different spin states. These states are fully stable, where the optical pumping provides \textit{Amplification} to fully compensate the still finite lifetime.

\section{Model}

To create the network of nodes we make use of the possibility to spatially pattern the potential $V(r)$ of exciton-polaritons \cite{20,21,22} as shown in Fig.1a and Fig.1b. The polaritons will be free to move in the shaded area, allowing their transport between nodes (circular confined area) through $1D$ channels.

We model the dynamics of the polariton field with right $\Psi_{+}(\textbf{r},t)$ and left $\Psi_{-}(\textbf{r},t)$ circular polarization, using the spinor Gross-Pitaevskii (GP) equation \cite{30,16},

\begin{equation}
\label{GP}
\begin{split}
i\hbar\frac{\partial\Psi_{+}}{\partial t}=(E_{LP}(\textbf{k})+V(\textbf{r})-\frac{i\hbar}{2\tau}+\alpha_{1}|\Psi_{+}|^2\\
 +\alpha_{2}|\Psi_{-}|^2)\Psi_{+}+\Delta(\textbf{r})\Psi_{-}+\mathfrak{F_{+}}(t,\textbf{r}), \\
i\hbar\frac{\partial\Psi_{-}}{\partial t}=(E_{LP}(\textbf{k})+V(\textbf{r})-\frac{i\hbar}{2\tau}+\alpha_{1}|\Psi_{-}|^2\\
 +\alpha_{2}|\Psi_{+}|^2)\Psi_{-}+\Delta(\textbf{r})\Psi_{+}+\mathfrak{F_{-}}(\textbf{r},t),
\end{split}
\end{equation}

$\tau$ is the finite lifetime of polaritons, $\alpha_{1}$ and $\alpha_{2}$ characterize the nonlinear interactions, $\mathfrak{F}_{s}(\textbf{r},t)$ represents the right ($s=+$) and left ($s=-$) circularly polarized components of the optical excitation and $E_{LP}(\textbf{k})$ represents the non-parabolic dispersion of the lowest polariton band\cite{5} (assuming a near-resonant excitation higher energy states are not populated).

\begin{figure}[h]
\includegraphics[width=8.7cm]{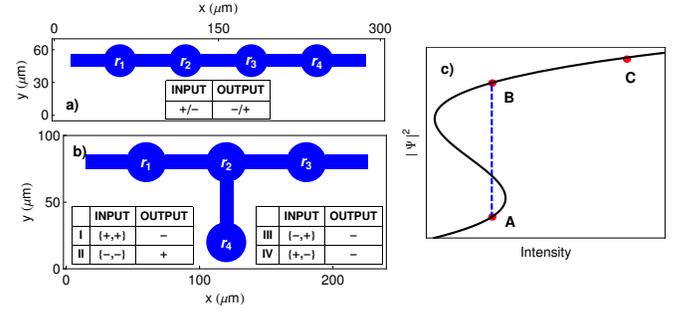}
\caption{Figure a) shows the confinement potential $V(\textbf{r})$ proposed for the NOT gate. The darker area corresponds to a potential of $V(\textbf{r})=0$, while the white area corresponds to $V(\textbf{r})=5meV$. Similarly, plot b) shows the potential used for the AND gate. Figure c) shows the bistablity at the nodes of the system. The dashed vertical line corresponds to the background intensity given by $|F_{s}|^{2}$. Points C and B correspond to the polariton density when a pulse $\mathcal{P}_{s}$ is applied and after it has finished, respectively.}
\label{Fig1}
\end{figure}

The polaritons are created with right ($\sigma^+$) and left ($\sigma^-$) circularly spin polarized components in each node using a normal incident near resonant Gaussian pump.
\begin{equation}
\label{F}
\mathfrak{F}_{s}(\textbf{r},t)=\sum_{i=1}^{4}F_{s}Exp[-\frac{(\textbf{r}-\textbf{r}_{i})^{2}}{\textit{l}^2}-\frac{i\varepsilon_{p}t}{\hbar}] +\mathcal{P}_{s}(\textbf{r},t),
\end{equation}
where $\textbf{r}_{i}$ is the center of each node (see Fig.\ref{Fig1}). In the channels, there is no optical pumping, such that exciton-polaritons are free to propagate ballistically.

The potential $\Delta(r)$, appears along the channels, as a consequence of the energy splitting of the optical modes with polarization parallel and perpendicular to the channel length \cite{25}. This splitting results in a similar energy splitting, of the exciton-polaritons \cite{26}, inversely proportional to the width of the channel. Such potential will induce the precession of the polarization along the channels since it mixes $\Psi_{+}$ and $\Psi_{-}$ states as can be appreciated in Eq.(\ref{GP}).

When the energy of the pump is chosen above $\sqrt{3}\hbar/(2 \tau)$, the dependence of the polariton density $|\Psi_{s}|^2$  on the pump power $|\mathfrak{F}_{s}|^2$ follows Fig.1c. Due to the presence of non-linear interactions the curve takes an S-shaped form, characteristic of bistability \cite{23, 24}. For the pump power indicated by the vertical dashed line the system can exist in one of two stable states marked by the points A and B.

Generally $|\alpha_{1}|>>|\alpha_{2}|$ \cite{13,Vladimirova}. Therefore, there are only weak interactions between polaritons with opposite spin components and one can expect a similar bistable response for each of the $\sigma^{+}$  and $\sigma^{-}$  components. With this spin degree of freedom, the system becomes multistable \cite{16} and one can write the input state using the circular polarization degree of the system,  $\rho=\frac{|\Psi_{+}|^2-|\Psi_{-}|^2}{|\Psi_{+}|^2+|\Psi_{-}|^2}$ as follows.

The first term in Eq.(\ref{F}) provides a Gaussian elliptically polarized cw pump which creates polaritons $\sigma^{+}$ and  $\sigma^{-}$ on the lower branch (around point A in Fig.1c). Sending a pulse $\mathcal{P}_{+}$, we can increase the intensity of the excitation for the $\sigma^{+}$  polarization, then the $\sigma^{+}$  state of polaritons is excited to the upper branch (point C). When the $\mathcal{P}_{+}$ pulse is turned off, the $\sigma^{+}$  state of polaritons moves to the point B. Meanwhile  $\sigma^{-}$  polaritons remain around point A during the whole process. In this case, the final state is strongly $\sigma^{+}$  polarized.

Notice that the cw pump is slightly biases to $\sigma^{+}$ so as to favor one polarization in the case where two of them meet (we will use this to construct AND gates shortly). Initially the population of $\sigma^{+}$ polaritons will be slightingly bigger than the $\sigma^{-}$ polaritons, but both will initialize in the lower bistable branch.

\section{Universal Logic gates}

\begin{figure}[h]
\includegraphics[width=8.cm]{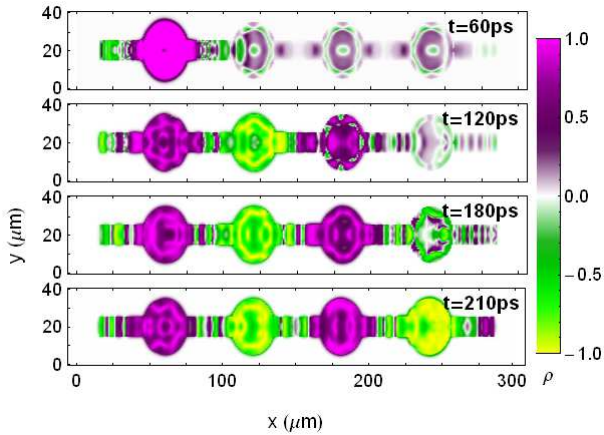}
\caption{Evolution of the circular polarization degree, $\rho$, of the system at different times. In the first row we can see the system at the time $t_0=60ps$, when the input pulse is sent writing the state of the first node. The pulse has a duration of $\tau=18 ps$, therefore for the successive times shown the system is evolving without the pulse. The state reached at $t=210ps$ is a stable state.}
\label{Fig2}
\end{figure}

In Fig.1a we show the proposed architecture for the NOT gate. The input signal is sent by directing the pulse over the first node (see top diagram in Fig.\ref{Fig2}), as
\begin{equation}
\label{PN}
\mathcal{P}_{+}^{N}(\textbf{r},t)=P_{+}^{N}Exp[-\frac{(\textbf{r}-\textbf{r}_{1})^{2}}{\textit{l}^2}-\frac{i\varepsilon_{p}t}{\hbar}-\frac{(t-t_{0})^2}{\delta t^{2}}].
\end{equation}
Once the input node is initialized, the pulse $\mathcal{P}$ can be turned off. The polaritons in the initialized node increase their potential energy (blueshift effect \cite{5}) and are accelerated towards the channels where they travel ballistically between nodes. The potential $\Delta(\textbf{r})$ causes a precession of the spin of polaritons travelling in the channel and it results in a complete inversion of the spin for the polaritons arriving at the next node.

When an incoming flux of polaritons arrives at an un-switched node, the increase in the density of polaritons induces a "jump" to the upper branch of the bistable curve (point B Fig.\ref{Fig1}c). This results in a switch of the node with a polarization opposite to the preceeding one. Notice that in this case the switching mechanism is different from the external pulse used to write the input state.

In other words, subsequent nodes switch with inverted spin polarization, as shown in Fig.2.
Each connection between nodes realizes a NOT logic gate with the input and output given by the truth table in the inset in Fig.1a. In this scheme there is complete \textit{Cascadability} and \textit{Amplification}.


%
\begin{figure*}[t]
\includegraphics[width=17cm]{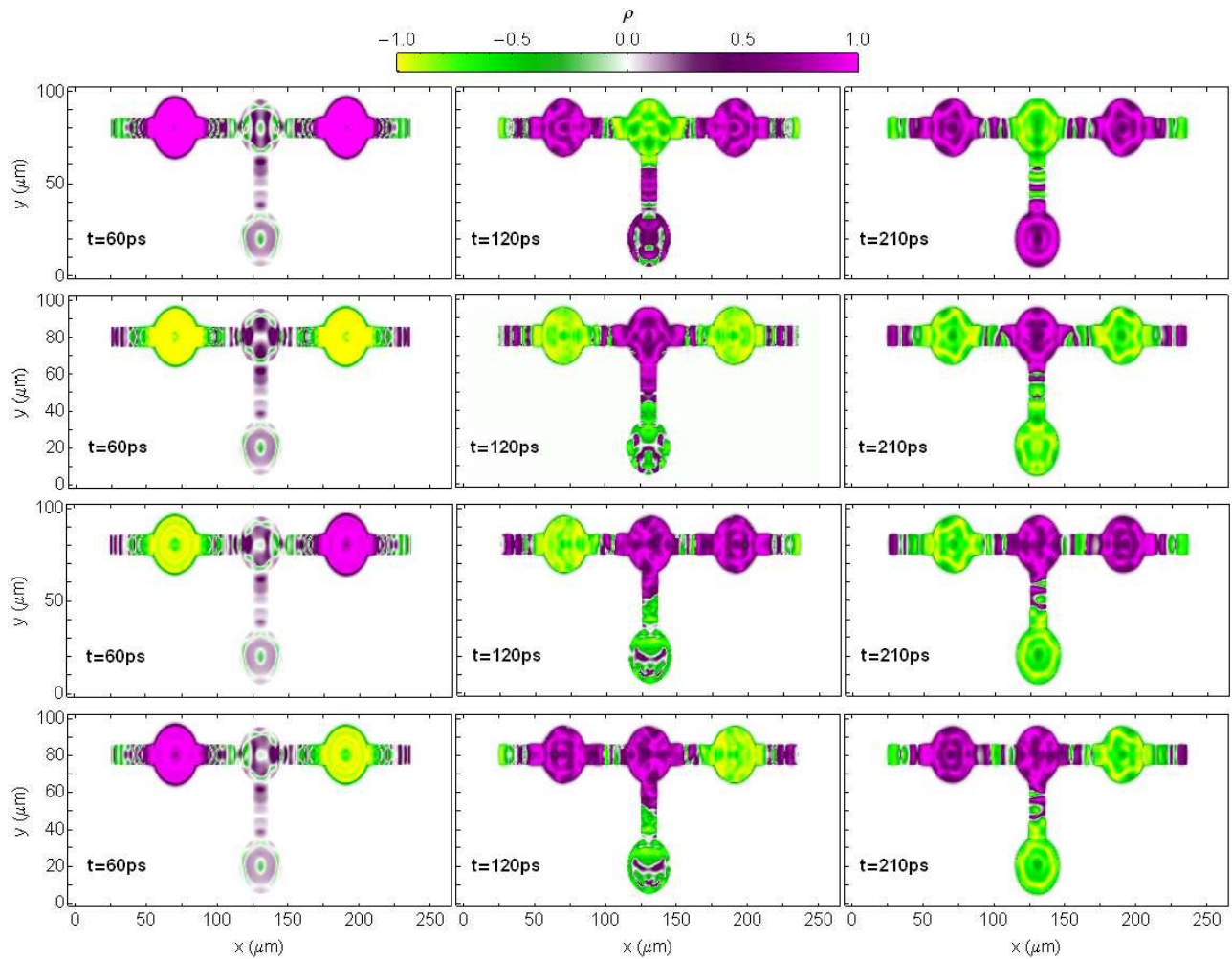}
\caption{We show the dynamics of the AND gate for different inputs; each row represents a different process where the input was changed according to the inset in Fig.1.b, i.e. row 1 is the process I, row 2, process II , etc. In the first column we show the time where the pulse is sent to write the state of the input nodes. For the successive times shown the system is evolving without the input pulse}
\label{Fig3}
\end{figure*}
To realize Universal Logic we also require an AND type logic gate with the truth table shown in the inset to Fig.1b. Here we will use the potential profile shown in Fig.1b. The left ($r_{1}$) and right ($r_{3}$) nodes are the input and can again be set with the application of laser pulses as,
\begin{equation}
\label{PA}
\begin{split}
\mathcal{P}_{s}^{A}(\textbf{r},t)=(P_{s}^{a}Exp[-\frac{(\textbf{r}-\textbf{r}_{1})^{2}}{\textit{l}^2}]+P_{s}^{b}Exp[-\frac{(\textbf{r}-\textbf{r}_{3})^{2}}{\textit{l}^2}])\\
*Exp[-\frac{i\varepsilon_{p}t}{\hbar}-\frac{(t-t_{0})^2}{\delta t^{2}}].
\end{split}
\end{equation}
By changing the values of the amplitudes $P_{s}^{a}$ and $P_{s}^{b}$, the state of the input nodes can be written as desired; for instance, the state on the first column of Fig.3 is achieved with $P_{-}^{a}=P_{-}^{b}=0$ and $P_{+}^{a}=P_{+}^{b}=0.5 meV\mu m^{-1}$.

Notice that due to the \textit{Cascadability} of the system, the input can also be set by connection to earlier parts of a larger circuit. The central node ($r_{2}$) can be considered as the output, the signal of which propagates to the lower node ($r_{4}$).

The time evolution of the AND gate for different inputs is shown in Fig.3. The state of the central node is determined by the state of the two input nodes.
Since the cw field is slightly elliptically polarized, the $\sigma^{+}$  polarization is favoured allowing the AND functionality.
Note that the change to elliptical polarization does not inhibit NOT gate operation such that the whole circuit is excited with a cw uniform polarization.
The typical switching time of the AND and NOT gates is of the order of $25-50ps$ per node corresponding to a single gate repetition rate in the range of tens of GHz. This figure of merit could be further improved by decreasing the distance between nodes or using cavities with a smaller polariton effective mass.

For the calculations we used the following parameters: $L=11\mu m$,$\tau=18 ps$, $\alpha_{2}/\alpha{1} =-0.22$, $l=10\mu m$, $\delta t=18ps$, $t_0=60ps$, $F_{-}=0.15meV\mu m^{-1}$,$F_{+}=0.17meV\mu m^{-1}$, $\varepsilon_{p}=0.5meV$ (energy with respect to the lowest energy of the polariton band) and $\Delta(\textbf{r})=0.09meV$, for $\textbf{r}$ inside the channel and $\Delta(\textbf{r})=0$, otherwise.

When tuning parameters, the appearance of bistability together with an inversion of the spin along the channels is crucial. For the first matter, the energy of the laser must be chosen above $\sqrt{3}\hbar/(2 \tau)$; while for the precession of the spin, it is fully determined by the value of the length of the channel and the polarization splitting $\Delta(\textbf{r})$ which is defined by the width of the channel \cite{25,26}. A narrower channel would increase the rotation of the spin and in order to obtain exactly the opposite spin, the length of the channel would have to be shorter. Therefore the system allows some freedom about the dimensions of the system as long as the inversion of the spin takes place between nodes.

In our proposal, it is also important that $\alpha_2$ is negative, such that interactions between opposite spin polarisations introduce a redshift of the energy. This prevents a given polarization switching to the high intensity state if the opposite polarization has already been switched in the same node, allowing input-output isolation (there is suppression of feedback to earlier nodes in the circuit). This is important for allowing a binary logic, where the state of each node takes only one of two possible values.

The power consumption of the device is expected to be dominated by the cw background field required for \textit{Amplification} of the signals. Polarization multistability was previously demonstrated\cite{27} with a power consumption of $50 W cm^{-2}$, but it can be further improved by increasing the life time of polaritons since, for switching, the polariton density injected must give a blueshift due to polariton-polariton interactions, determined by $\varepsilon_{p}=|\Psi_{s}|^2\alpha_{1}\sim\sqrt{3}\hbar/(2 \tau)$. Here, the polariton density, scales with the pump power, roughly as $|\Psi_{s}|^2\sim F_{s}^2\tau^2$. Thus, the required pump power for bistability scales as $F_{s}^2\sim1/(\alpha_{1}\tau^3)$. In other words an improvement of the polariton lifetime by a factor of $20$ (as reported recently\cite{17}) can be expected to result in a power consumption an order of magnitude lower than the power consumptions in state-of-the-art photonic crystal cavities \cite{Nozaki}.

\section{Conclusions}

We have presented a scheme for the realization of complete photonic integrated circuits in a single compact solid-state system. Unlike previous schemes, we demonstrate the criteria required for fully functional devices including complete \textit{Universal Logic}. By making use of spin-multistability there is total logic-level \textit{Amplification} of the states, completely compensating the finite lifetime of particles in the system. We expect these results to open a new horizon for the development of novel optical based computers.  A further interesting perspective is the ability to engineer polariton potentials optically \cite{28} allowing one to dynamically update the circuit design. This could allow the completion of long circuits in small areas, giving a route for an optical device to compete with electronics in terms of device size despite the comparably long wavelength of light. Since any transition between optical and electrical devices would be most likely to occur in stages, hybrid electro-optic systems are also of great interest. The presented scheme can be expected to be controlled by the application of electric fields, which shift the polariton energy via the quantum confined Stark effect \cite{29}.

We Acknowledge Prof. I. A. Shelykh for valuable discussions.



\begin{thebibliography}{99}


\bibitem{1}Q. F. Xu, B. Schmidt, S. Pradhan and M. Lipson, Nature, \textbf{435}, 325 (2005).

\bibitem{2}M.Liu, X. Yin, E. Ulin-Avila, B. Geng, T. Zentgraf, L. Ju, F. Wang and X. Zhang, Nature, \textbf{474}, 64 (2011).

\bibitem{3}G. Grosso, J. Graves,A. T. Hammack, A. A. High, L. V. Butov, M. Hanson, A. C. Gossard, Nature Phot., \textbf{3}, 577 (2009).

\bibitem{4}L. Fan, J. Wang, L. T. Varghese, H. Shen, B. Niu, Y. Xuan, A. M. Weiner and M. H. Qi, Science, \textbf{335}, 447 (2012).

\bibitem{5} A. V. Kavokin, J. J. Baumberg, G. Malpuech and F. P. Laussy, Oxford University Press (2007).

\bibitem{6} D. A. B. Miller, Nature Photon., \textbf{4}, 3 (2010).

\bibitem{7} R. W. Keyes, Science, \textbf{230}, 138 (1985).

\bibitem{8} D. Sanvitto, S. Pigeon, A. Amo, D. Ballarini, M. De Giorgi, I. Carusotto, R. Hivet, F. Pisanello, V. G. Sala, P. S. S. Guimaraes, R. Houdre, E. Giacobino, C. Ciuti, A. Bramati and G. Gigli, Nature Photon., \textbf{5}, 610 (2011).

\bibitem{9} T. Gao, P. S. Eldridge, T. C. H. Liew, S. I. Tsintzos, G. Stavrinidis, G. Deligeorgis, Z. Hatzopoulos and P. G. Savvidis, Phys. Rev. B, \textbf{85}, 235102 (2012).

\bibitem{10}G. Christmann, C. Coulson, J. J. Baumberg, N. T. Pelekanos, Z. Hatzopoulos, S. I. Tsintzos and P. G. Savvidis, Phys. Rev. B, \textbf{82}, 113308 (2010).

\bibitem{11} E. Wertz, A. Amo, D. D. Solnyshkov, L. Ferrier, T. C. H. Liew, D. Sanvitto, P. Senellart, I. Sagnes, A. Lemaitre, A. V. Kavokin, G. Malpuech and J. Bloch, Phys. Rev. Lett., \textbf{109}, 216404 (2012).

\bibitem{12} I. A. Shelykh, Y. G. Rubo, A. V. Kavokin, T. C. H. Liew and G. Malpuech, Semicond. Sci. Technol., \textbf{25}, 013001 (2010).

\bibitem{13} C. Leyder, T. C. H. Liew, A. V. Kavokin, I. A. Shelykh, M. Romanelli, J. Ph. Karr, E. Giacobino, and A. Bramati, Phys. Rev. Lett., \textbf{99}, 196402 (2007).

\bibitem{14} A. Amo, T. C. H. Liew, C. Adrados, R. Houdre, E. Giacobino, A. V. Kavokin, and A. Bramati, Nature Photon., \textbf{5}, 610 (2011).

\bibitem{Dario}D. Ballarini, M. De Giorgi, E. Cancellieri, R. Houdré, E. Giacobino, R. Cingolani, A. Bramati, G. Gigli, D. Sanvitto, arXiv:1201.4071 (2012),accepted for publication in Nature Comm. (2013).

\bibitem{Giorgi} M. De Giorgi, et al., Phys. Rev. Lett., \textbf{109}, 266407 (2012)

\bibitem{15} T. C. H. Liew, A. V. Kavokin, I. A. Shelykh, Phys. Rev. Lett., \textbf{101}, 016402 (2008).	

\bibitem{16} N. A. Gippius, I. A. Shelykh, D. D. Solnyshkov, S. S. Gavrilov, Y. G. Rubo, A. V. Kavokin, S. G. Tikhodeev and G. Malpuech,  Phys. Rev. Lett., \textbf{98}, 236401 (2007).

\bibitem{17} B. Nelson, G. Liu, M. Steger, D. W. Snoke, R. Balili, K. West and L. Pfeiffer, arXiv:1209.4573 (2012).

\bibitem{18} E. Kammann, T. C. H. Liew, H. Ohadi, P. Cilibrizzi, P. Tsotsis, Z. Hatzopoulos, P. G. Savvidis, A. V. Kavokin and P. G. Lagoudakis, Phys. Rev. Lett., \textbf{109}, 036404 (2012).

\bibitem{19} I. A. Shelykh, R. Johne, D. D. Solnyshkov and G. Malpuech, Phys. Rev. B, \textbf{82}, 153303 (2010).

\bibitem{20} R. I. Kaitouni, O. El Daif, A. Baas, M. Richard, T. Paraiso, P. Lugan, T. Guillet,
F. Morier-Genoud, J. D. Ganiere, J. L. Staehli, V. Savona and B. Deveaud,  Phys.
Rev. B \textbf{74}, 155311 (2006).


\bibitem{21}R. B. Balili, D. W. Snoke, L. Pfeiffer and K. West, Appl.Phys. Lett. \textbf{88}, 031110 (2006).


\bibitem{22} C. W. Lai, N. Y. Kim, S. Utsunomiya, G. Roumpos, H. Deng, M. D. Fraser, T. Byrnes, P. Recher, N. Kumada, T. Fujisawa and Y. Yamamoto, Nature (London) \textbf{450}, 529
(2007).

\bibitem{30} I. Carusotto  and C. Ciuti, Phys. Rev. Lett. \textbf{93}, 166401 (2004).

\bibitem{25} A. Kuther, M. Bayer, T. Gutbrod, A. Forchel, P. A. Knipp, T. L. Reinecke, R. Werner, Phys. Rev. B, \textbf{58}, 15744 (1998).

\bibitem{26} G. Dasbach, C. Diederichs, J. Tignon, C. Ciuti, P. Roussignol, C. Delalande, M. Bayer and A. Forchel, Phys. Rev. B, \textbf{71}, 161308(R) (2005).

\bibitem{24} A. Baas, J. Ph. Karr, H. Eleuch and E. Giacobino, Phys. Rev. A \textbf{69}, 023809 (2004).

\bibitem{23}D. M. Whittaker, Phys. Rev. B 71, 115301 (2005).

\bibitem{Vladimirova} M. Vladimirova,  S. Cronenberger, and D. Scalbert, K.V. Kavokin, A. Miard, A. Lemaitre, J. Bloch, D. Solnyshkov, G. Malpuech, A. V. Kavokin,Phys. Rev. B \textbf{82}, 075301 (2010).

\bibitem{27} T. K. Paraiso, M.  Wouters, Y. Leger, F. Morier-Genoud and B. Deveaud-Pledran, Nature Materials, \textbf{9}, 655 (2010).

\bibitem{Nozaki} K. Nozaki, T. Tanabe, A. Shinya, S. Matsuo, T. Sato, H. Taniyama, and M. Notomi, Nature Phot.,\textbf{4}, 477 (2010).

\bibitem{28} A. Amo, S. Pigeon, C. Adrados, R. Houdre, E. Giacobino, C. Ciuti and A. Bramati, Phys. Rev. B \textbf{82}, 081301(R) (2010).

\bibitem{29} T. C. H. Liew, A. V. Kavokin, T. Ostatnicky, M. Kaliteevski, I. A. Shelykh and R. A. Abram, Phys. Rev. B, \textbf{82}, 033302 (2010).



\end{thebibliography}
\end{document}